# Visualizing a Terahertz Superfluid Plasmon in a Two-Dimensional Superconductor

A. von Hoegen[1], T. Tai[1], C. Allington[1], M. Yeung[1], J. Pettine[1], M. Michael[3,4], E. Viñas Boström[3], X. Cui[2], K. Torres[2], A. E. Kossak[1], B. Lee[1], G. S. D. Beach[1], G. Gu[6], A. Rubio[3,5], P. Kim[2], N. Gedik[1]

[1]*Massachusetts Institute of Technology, 02139 Cambridge, United States*
[2]*Harvard University, Cambridge, Massachusetts 02138, USA*
[3]*Max Planck for the Structure and Dynamics of Matter, 22607 Hamburg, Germany*
[4]*Max Planck Institute for the Physics of Complex Systems, Nöthnitzer Str. 38, 01187 Dresden*
[5]*Center for Computational Quantum Physics (CCQ), Flatiron Institute, 10010 New York, United States*
[6]*Department of Condensed Matter Physics and Materials Science, Brookhaven National Laboratory, Upton, NY 11973, USA.*

**The superconducting gap defines the fundamental energy scale for the emergence of dissipationless transport and collective phenomena in a superconductor[1-3]. In layered high-temperature cuprate superconductors, in which the Cooper pairs are confined to weakly coupled two-dimensional copper-oxygen planes[4,5], terahertz (THz) spectroscopy at sub-gap millielectronvolt energies has provided crucial insights into the collective superfluid response perpendicular to the superconducting layers[6-9]. However, within the copper-oxygen planes the collective superfluid response manifests as plasmonic charge oscillations at energies far exceeding the superconducting gap, obscured by strong dissipation[2,6,9,10]. Here, we present spectroscopic evidence of a below-gap, two-dimensional superfluid plasmon in few-layer $Bi_2Sr_2CaCu_2O_{8+x}$ and spatially resolve its deeply sub-diffractive THz electrodynamics. By placing the superconductor in the near-field of a spintronic THz emitter, we reveal this distinct resonance—absent in bulk samples and observed only in the superconducting phase—and determine**

**its plasmonic nature by mapping the geometric anisotropy and dispersion. Crucially, these measurements offer a direct view of the momentum- and frequency dependent superconducting transition in two dimensions. These results establish a new platform for investigating superfluid phenomena at finite momenta and THz frequencies, highlighting the potential to engineer and visualize superconducting devices operating at ultrafast THz rates.**

## Introduction

The superconducting ground state is described by a condensate wavefunction $\psi = \Delta e^{i\chi}$ and its dynamics are perturbative excitations of the amplitude $\Delta$ and phase $\chi$, commonly referred to as Higgs and Nambu-Goldstone modes, respectively[11-13]. In the long-wavelength limit (q ~ 0), the Higgs mode coincides with the superconducting energy gap $h\nu_H = 2\Delta$[14]. Located at the lower bound of the quasiparticle continuum, the decay of the Higgs mode into single-particle excitations is strongly kinematically suppressed[14]. Experimentally, under nonlinear terahertz (THz) excitation, the Higgs mode appears as coherent oscillations in conventional BCS superconductors and as an overdamped, driven response in high-temperature cuprate superconductors[1,15,16]. Linear coupling between the Higgs mode and electromagnetic fields is symmetry forbidden[1,13].

In contrast, phase fluctuations are directly influenced by long-range Coulomb interaction[17], which lift the Goldstone mode far above the superconducting gap energy (Anderson-Higgs mechanism), producing a gapped plasma oscillation at $\Omega_P$, akin to the usual plasma oscillations in normal metals[18,19]. In conventional superconductors, plasma oscillations typically occur at eV energy scales due to the high superfluid stiffness $\rho_s$, which determines the energy cost of imposing a phase gradient[6]. In this high-energy regime, the plasmonic character of the superfluid is obscured by strong dissipation and quasiparticle pair breaking[2,6,10].

In highly anisotropic superconductors, such as the high-temperature cuprates, the superfluid is confined to two-dimensional (2D) copper–oxygen ($CuO_2$) planes that are only weakly coupled, resulting in a significantly reduced superfluid stiffness along the out-of-plane direction[4,6]. This gives rise to a below-gap superfluid plasmon, polarized perpendicular to the $CuO_2$ planes, associated with Josephson tunneling of Cooper pairs[7,8]. These coherent plasma-like oscillations of the superfluid, known as Josephson plasmons, are routinely observed with linear THz spectroscopy at the relevant sub-gap millielectronvolt (meV) energy scales[7,8]. In contrast, within each $CuO_2$ plane, the superfluid stiffness is much stronger, and the collective response only occurs at energies far exceeding the superconducting gap, where it is mostly incoherent[2,6,9,10].

The superfluid character of this plasmon can be recovered by reducing the thickness of the superconductor toward the 2D limit[20,21]. In particular, a dimensional crossover from 3D to 2D plasmonic behavior takes place once the in-plane momentum $q$ becomes comparable to the film thickness, $qd \sim 1$. In this regime, hybridization of the surface plasma modes gives rise to a low-energy dispersive branch ($\Omega \sim \sqrt{q}$)[20,22] (see Supplementary Section S20), leading to the emergence of an in-plane superfluid mode with strong plasmonic character at meV energies[22]. As a result, this in-plane superfluid plasmon acquires additional momentum as the dispersion is shifted away from the light line, rendering it inaccessible to conventional far-field THz experiments[23].

To access this finite-momentum regime, THz radiation ($\lambda$ ~ 300 µm) can be confined to deeply sub-wavelength dimensions at the apex of nanoscopic metallic tips (a ~ 10 nm)[24-26], providing a broad momentum distribution, coupled to metallic structures with a well-defined spatial periodicity[27,28], or generated and detected within the near-field region of the sample[29-32]. However, tip-based measurements provide limited coupling to in-plane modes owing to their primarily out-of-plane sensitivity, due to the strong vertical confinement at the emitter apex[24,27,33]. Other methods, such as gratings, are challenging to implement due to the reduced lateral size of exfoliated single crystal samples[34,35]. Overcoming these challenges would reveal critical information about the ability to tailor 2D superconductivity on ultrafast THz timescales and highly confined spatial scales, providing a more sensitive probe of phase fluctuations[36].

In this paper, we take advantage of the quasi-2D nature of the high-temperature superconductor $Bi_2Sr_2CaCu_2O_{8+x}$ (BSCCO) and utilize deeply sub-diffractive THz hyperspectral imaging to directly visualize a 2D in-plane superfluid plasmon. In particular, we interface a few-layer single crystal of BSCCO—which retains its superconducting state down to the monolayer limit[34]—with a spintronic THz source and locally trigger THz emission with an ultrashort laser pulse[29,37]. This confined THz source launches an in-plane superfluid plasmonic wave, which radiates into the far field due to the finite size of the sample. The plasmonic mode softens and disappears close to the critical temperature of BSCCO, clarifying its superfluid character. By spatially scanning the confined THz source, we capture a hyperspectral, phase-resolved visualization of the 2D superfluid plasmon. From these spatially resolved measurements, we determine the in-plane plasmon dispersion and provide direct evidence of a frequency- and momentum-dependent superconducting transition in two dimensions.

### *THz Micro-Spectroscopy of few-layer $Bi_2Sr_2CaCu_2O_{8+x}$*

Figure 1a shows an optical microscopy image of a mechanically exfoliated sample (28 nm thickness, see Figure 1b) of optimally doped $Bi_2Sr_2CaCu_2O_{8+x}$ placed on a spintronic THz emitter[29,37] using a dry-transfer method (see Supplementary Section S3). To prevent degradation, the sample is encapsulated with 30-nm thick hBN and kept under vacuum and at low temperatures (< 150 K).

To measure the local THz light–matter coupling in BSCCO, we use tightly focused ultrashort near-infrared laser pulses to locally excite THz emission from the spintronic emitter (see Figure 1c). The resulting THz radiation is confined to a deeply sub-diffractive lateral size of 15 μm and immediately interacts with the sample before diverging into the far-field (see Supplementary Section S1 and S10). Crucially, to avoid photoexcitation, we isolate the sample from the residual near-infrared laser via a distributed Bragg reflector (DBR) (see Supplementary Section S2). After refocusing the transmitted THz radiation into a (110) cut ZnTe crystal, we record the THz waveform by standard electro-optical sampling with an ultrashort gate pulse (see Supplementary Section S1).

Figure 1d shows the measured time-resolved THz electric field when transmitted through the BSCCO sample (top) in the superconducting state (T = 10 K), and a reference of the unaffected field measured away from the sample (bottom). The field transmitted through the sample is strongly attenuated, stretched and exhibits low-frequency coherent oscillations following the main pulse, which are absent in the reference field. Figure 1e shows the corresponding amplitude spectrum obtained by fast Fourier transform (FFT) of the time-resolved electric field, displaying a non-uniform attenuation with pronounced minimum near 1.4 THz. By referencing the complex amplitude spectrum of the sample to that of the reference field, we can extract the phase and amplitude of the complex transmission coefficient, shown in Figures 1f and g, respectively. The absolute minimum in the transmission amplitude at 1.4 THz is accompanied by a rapid phase shift from π/2 at lower frequencies to -π/2 at higher frequencies—consistent with the resonant absorption of a collective mode.

***Effective THz Conductivity***

To analyze these results, we consider a model based on Fresnel coefficients in the thin film limit to extract the effective frequency-dependent THz conductivity of our sample[38] (see Supplementary Sections S12 and S17). Figures 2a and b show the result of this analysis in the form of the real Re($\sigma$) and imaginary Im($\sigma$) parts of the effective THz conductivity, respectively. As expected, the real part of the conductivity, which is associated with dissipation, comprises a narrow peak at 1.4 THz with a Lorentzian line shape[39] (see Figure 2a). The corresponding imaginary part, which encodes the phase relation between the incident THz field and the induced transient currents within the sample, shows a crossover from negative to positive values (see Figure 2b). These are the hallmark characteristics of a collective mode within a Lorentz oscillator model and are strikingly different from the known THz response of bulk BSCCO[6]. In a bulk superconductor, we expect a $1/\omega$ divergence of the imaginary conductivity due to the inductive inertia of the superfluid and a small dissipative response around $\omega = 0$ reflecting the uncondensed electrons[39], also plotted on the same scale in Figures 2a and b as dashed lines[40]. This departure from the conventional behavior is attributed to the sample's finite size, consistent with a nonlocal, finite-momentum response[22,25,26].

We further investigated the nature of this collective mode as a function of temperature. Figures 2c and d show the THz conductivity measured at 95 K, just above the critical temperature ($T_c$ = 87 K) of BSCCO. At this temperature, the width of the resonance is substantially increased, indicative of the higher dissipation in the normal state of BSCCO compared to the low-temperature superconducting phase.

Figures 2e and f provide a comprehensive measurement of Re($\sigma$) and Im($\sigma$) as a function of temperature and frequency, revealing that the resonance abruptly vanishes within a 10 K window around $T_c$. The inset of Figure 2e shows the frequency $\Omega_R$ and the width $\gamma_R$ of the resonance. We find that the transition is linked to a decrease of $\Omega_R$ and a significant increase of $\gamma_R$, which is associated with the damping rate.

We analyzed the measured complex conductivity with a Drude-Smith-Lorentz[39,41] model, which captures both the free-carrier response in spatially confined systems and the resonance (see Supplementary Section S12). The full results of this analysis are shown in Supplementary Data Figures S16-20. Besides the resonance frequency $\Omega_R$ and damping rate $\gamma_R$, this analysis yields the respective amplitudes $A \propto \Omega_{\mathrm{P}}^2$ of the free carrier response and the resonance. Here, $\Omega_{\mathrm{P}}$ denotes the plasma frequency of either component (see Supplementary Section S12). This allows us to estimate the spectral weight $S_\sigma$ contained within the real part of the optical conductivity, which in bulk systems satisfies the frequency-sum (f-sum) rule $S_\sigma = \int_0^\infty \mathrm{Re}(\sigma)\,\mathrm{d}\omega \propto n \propto \sum \Omega_{\mathrm{P}}^2$[42], where $S_\sigma$ is proportional to the total charge carrier density $n$. Our data suggests that $S_\sigma$ remains largely conserved across all temperatures (see Supplementary Data Figure S20) and the corresponding individual contributions of $\Omega^2_{\mathrm{P,\,free}}$ and $\Omega^2_{\mathrm{P,\,mode}}$ show a clear crossover from high temperatures to low temperatures associated with a spectral weight transfer from $\Omega^2_{\mathrm{P,\,free}}$ to $\Omega^2_{\mathrm{P,\,mode}}$ (see Inset in Figure 2f). This implies that the free charge carriers present at high temperatures form a bound electronic mode at low temperatures due to the reduced scattering in the superconducting state.

Finally, we note that while providing qualitative insight, this analysis is generally applicable only to far-field optics in the long-wavelength limit (q = 0). For a more accurate description, we must take into account the deeply sub-diffractive electrodynamics of our experiment[43]. For this, we consider the electromagnetic wave equation in the near-field limit. We find that the electric field emitted into the far field $E_{\mathrm{emit}}$ is dominated by nonlocal dynamics within the BSCCO thin film (see Supplementary Section S17), which yields an effective response function

$$\frac{E_{\mathrm{emit}}(\omega, y') - E_0(\omega)}{E_0(\omega)} = T(\omega, y') - 1 = \frac{\mu_0 c^2}{\varepsilon_r \omega^2 L}\left[\partial_y \sigma(\omega, y', y = L) - \partial_y \sigma(\omega, y', y = 0)\right].$$

Here, $\varepsilon_r$ is the effective dielectric constant of the environment, $y'$ is the excitation coordinate with respect to the sample boundaries $y = [0, L]$ and $\sigma$ is the complex conductivity of BSCCO. This equation implies that the measured local response function, $T(\omega, y')$, holographically encodes the dynamics of the boundary charge density. These charge dynamics arise from propagating plasmonic modes inside the sample launched within the near-field of the confined THz source $E_0$ and made resonant when their wavelength matches the sample dimensions.

We note that while this analysis accurately reflects the experimental response (see Supplementary Section S17), a full quantitative characterization of the conductivity requires a theoretical framework that resolves the complete three-dimensional electromagnetic boundary conditions.

***THz Hyperspectral Imaging***

Next, to visualize these local THz electrodynamics, we scan the pump beam across the sample and record a complete THz waveform at each position (see Figure 3a). This approach allows us to record a THz hyperspectral image of the relative transmission amplitude and phase—referenced to the bare emitter—with subwavelength resolution, as shown in Figures 3b and c for a base temperature of 10 K. At frequencies above the resonance (2.5 THz), the shape of the attenuated THz transmission amplitude agrees well with the outline of the sample (dashed line in Figure 3b). At frequencies on (1.5 THz) and below (0.75 THz) resonance, the relative transmission amplitude develops a more complex shape which depends on the geometry of the sample. These geometric effects are especially pronounced at lower frequencies (0.75 THz), giving rise to a relative transmission amplitude larger than unity at the sharpest edges of the sample. These enhanced local fields likely arise from plasmonic channeling of THz radiation through hotspots at the sharpest spatial features of the sample[44,45](see Supplementary Section S23).

The corresponding phase of the complex transmission coefficient (see Figure 3c) is nearly uniform—with opposite signs—above and below resonance, while developing a wave-like pattern for frequencies close to the resonance.

For a more quantitative analysis of the spatial confinement of this mode, we fabricated a second sample with a more well-defined, almost rectangular shape with an aspect ratio of 3:2 (see Figure 3d). By virtue of a small external magnetic field $B_{ext}$ ~ 20 mT, oriented in plane (see Figure 3a), we can align the THz polarization with the two major axes of this sample (see Supplementary Section S8). Figure 3e shows the THz conductivity for these two configurations, denoted as vertical and horizontal, corresponding to the THz polarization aligned with the short and long sides of the sample, respectively. These measurements reveal a clear anisotropy and a pronounced redshift of the resonance when the THz field aligns with the long side of the rectangle.

Figures 3f and g show spatial maps of the spectral phase at resonance for the two orthogonal THz polarizations. Again, we observe a wave-like structure in the spatial maps, aligned with the polarization direction. This is further illustrated in Figures 3h and 3i, which show vertical and horizontal line cuts through the spatial patterns. By determining the width of the central half-cycle (red circles), we estimate the wavelength $\lambda_R$ from the full width at half maximum, and thus the momentum of the resonance, which in both cases agree well with the corresponding sample dimensions. This longitudinal momentum is directly provided by the broad momentum distribution in the near-field of the deeply sub-diffractive spintronic THz source.

Finally, we tracked the temperature dependence of the mode's resonance frequency and its spatial wavelength (see Supplementary Section S14). As before, the resonance frequency softens with increasing temperature and, near $T_c$, approaches the low-frequency cutoff of our experiment for both polarizations (see Figure 3j). To determine the temperature dependence of the wavelength $\lambda_R$, we extracted line cuts like the ones shown in Figures 3h and 3i for a set of temperatures up to 90 K at the corresponding $\Omega_R(T)$ (see Supplementary Data Figure S24 and S25). We find that the wavelength $\lambda_R$ and therefore the momentum remains nearly constant for all measured temperatures (see Figure 3k).

Based on all these experimental signatures, we assign the resonance that arises in the superconducting phase of BSCCO to a two-dimensional plasmonic excitation of the superfluid.

***Superfluid Plasmon***

The plasmonic response is intimately related to the dimensionality of the system, scaling as $\Omega_R \propto \sqrt{q}$[20] in the ultrathin film limit. This significantly enhances the plasmonic character of the mode in the few-THz frequency range (small $q$) with strong deviations from the light line, corresponding to a decreased group velocity, and enhanced light field confinement[46]. The relevant two-dimensional carrier density is given by $n_{2D} = n_{3D}d$, where $d$ is the thickness of the BSCCO flake, leading to the expression[20-22]

$$\Omega_R \propto \sqrt{n_{3D}d \cdot q}. \text{ (Eq.1)}$$

Therefore, normalizing the observed plasma frequencies by the square root of the sample thickness $\Omega_R/\sqrt{d}$ should reveal a monotonic trend characteristic of the specific material. With these considerations in mind, Figure 3l shows this quantity obtained from four distinct samples with varying thickness. All data points follow a universal scaling (dashed line), which is clearly distinct from the expected square-root dependence on momentum ($\sqrt{q}$). This discrepancy arises due to screening of the plasmon by the metallic spintronic emitter, resulting in an almost linear dispersion, $\Omega_R \propto q^\alpha$ with $\alpha \approx 1$ (see Supplementary Section S18), which agrees well with the data shown in Figure 3l (dashed line). From the slope of this linear dispersion, we can determine the phase velocity of the plasmon which is related to the 3D plasma frequency and the geometrical factors as $v_p = \Omega_{P,3D}\sqrt{\frac{td}{\varepsilon_r}}$ (see Supplementary Section S18). Here, $\varepsilon_r$ is the relative permittivity of the DBR and t is its thickness. Using this expression to fit the dispersion, shown in Figure 3l, we find $v_p = 0.099 \pm 0.003\ c$, which matches closely with the phase velocity of $0.11\ c$ calculated from literature[42].

Finally, following eq. 1, we can interpret the pronounced softening of $\Omega_R$ (see Figure 3j) as evidence for a reduction in the charge carrier density that forms the plasmonic resonance ($\Omega_R \propto \sqrt{n}$). At the lowest temperature in the experiment (T ≈ 10 K)—when most charge carriers are condensed—we can associate this density with the superfluid density, $n = n_s$. To substantiate this interpretation, we compare the previously measured temperature dependence of the superfluid density $n_s$ in BSCCO[40,47] to the full temperature dependence $\Omega_R$ (see Figure 3j) and find nearly excellent agreement, with only slight deviations at low temperatures. This behavior suggests that the plasmonic resonance reflects the condensed fraction of charge carriers, identifying it as a finite-momentum 2D superfluid plasmon.

Based on this conclusion, we estimate the in-plane superfluid stiffness $\rho_s$ of our 2D system. The superfluid stiffness measures the energy cost to impose a phase gradient and is related to the superfluid density by $\rho_s = \frac{\hbar^2 n_s}{4m^*}$ for parabolic bands with an effective mass $m^*$. Our measurement of the plasmon frequency $\Omega_R \propto \sqrt{\frac{n_s}{m^*}}$, therefore allows us to calculate the low-temperature limit of the superfluid stiffness at T = 10 K to be $\rho_s = 7.65 \pm 0.48$ meV which corresponds to $88.8 \pm 5.6$ K ~ $T_c$. From the data in Figure 3j, we can further extract the full temperature dependence of $\rho_s$, which is plotted in Figure 4a. As expected from previous measurements[40,48], the superfluid stiffness decreases more rapidly above T ~ 65 K, once it becomes comparable to thermal energy scale $k_B T$ (dashed line), due to the proliferation of vortex-antivortex pairs[48-50]. Crucially, this temperature scale coincides well with the onset of increased dissipation of the superfluid plasmon (see Figure 4b and Supplementary Data Figure S28). This abrupt increase in dissipation at T ≈ 60 K is unlikely to arise from closing of the superconducting gap, which is only weakly temperature dependent below 80 K[51]. To further substantiate these findings, we carried out a similar analysis on the temperature-dependent measurements taken on the samples with different dimensions and find a monotonic decrease of this onset temperature $T_{on}$ when plotted against the corresponding momentum (see Figure 4c).

To explain this scale dependence, we compare the energy stored within a smooth phase gradient to the self-energy of an unbound vortex–antivortex pair[52] (see Supplementary Section S16)—i.e. we ask when vortex unbinding becomes energetically more favorable than maintaining the smooth phase gradient imposed by the plasmon. From these considerations, we find that the highest momentum the condensate can support is inversely proportional to the Pearl length, $q_{\text{max}} \sim \frac{1}{\Lambda_P} \sim \rho_s$, which defines the length scale over which a vortex-antivortex pair is still logarithmically bound in a thin superconductor[53]. In other words, the superfluid stiffness defines the maximum permissible phase-gradient before the superfluid plasmon dissociates into unbound vortex-antivortex pairs and strong damping sets in. With this in mind, we expect the onset temperature to increase linearly as $q \rightarrow 0$, recovering the critical temperature $T_c$ (see Figure 4c, red square) in the long wavelength limit ($q = 0$), which agrees well with a linear extrapolation of our data (dashed line). Therefore, these results suggest a scale dependence of the dynamic Kosterlitz-Thouless-Berezinskii (KTB)-like transition[48], above which vortex-antivortex unbinding melts finite-momentum phase distortions of the superfluid.

***Conclusion***

To conclude, we presented THz spectroscopy measurements of optimally doped $Bi_2Sr_2CaCu_2O_{8+x}$ with deep sub-wavelength resolution, achieved by interfacing a few-layer flake with a spintronic THz source. In the superconducting phase, the THz light-matter coupling is dominated by a 2D superfluid plasmon, and temperature-dependent measurements revealed that both its frequency and amplitude track the superfluid density. Polarization-sensitive imaging captured the anisotropy of this plasmonic response, which is dominated by geometric confinement. Finally, by launching coherent superfluid plasma waves across macroscopic distances, we access a previously unexplored regime of two-dimensional superconductivity and observe a frequency- and momentum-dependent onset of dissipationless behavior in two dimensions. This ability to launch and detect superfluid plasmons provides a new scalable platform to engineer and visualize superconducting devices operating at THz frequencies[54].

In the future, similar measurements can be performed across the entire phase diagram of $Bi_2Sr_2CaCu_2O_{8+x}$ to provide new insights into the residual superconducting coherence in the normal state of under- and overdoped cuprates[36]. Measurements at sub-picosecond timescales and momenta outside the light cone would provide complementary information to time- and momentum-integrated probes such as the superconducting Nernst effect[55] and open new avenues for investigating intrinsic 2D superconductivity in ultra-clean single-crystal samples[56,57].

Finally, the experimental strategy presented above can straightforwardly be extended to all layered van der Waals materials and devices. These mechanically stacked van der Waals heterostructures or moiré materials, designed by a rotational or lattice mismatch of atomically thin crystals[58], give rise to strongly interacting electronic phases with characteristic low-energy collective excitations. Future progress in understanding these characteristic features requires new probes at their natural meV-energy scales, with µeV-energy resolution and sub-diffractive spatial resolution. The spintronic device infrastructure presented here is compatible with the current state-of-the-art device fabrication[56,57,59] and THz light-matter coupling remains strong in the 2D limit, giving rise to pronounced features in the spatially resolved THz transmission. Its inherent ultrafast nature further makes this method suitable to study the non-equilibrium physics of these strongly correlated 2D quantum systems, when subjected to strong photoexcitation.

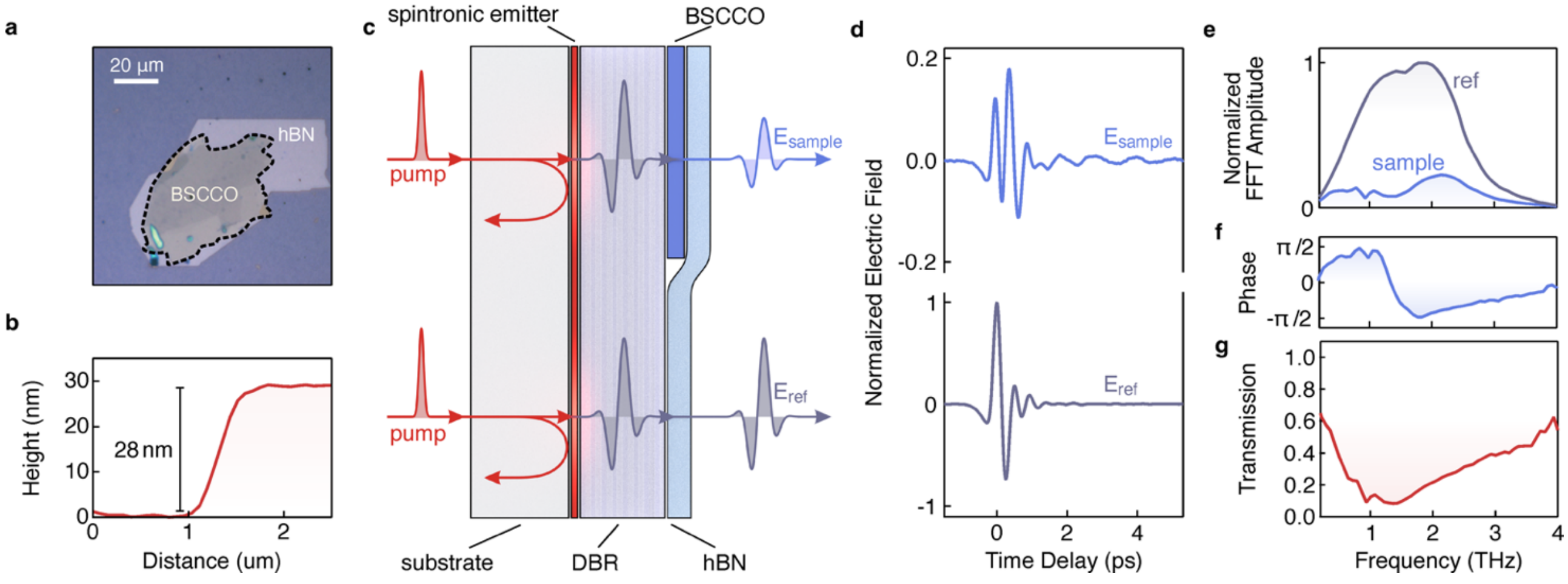

**Figure 1: THz micro-spectroscopy of few-layer $Bi_2Sr_2CaCu_2O_{8+x}$. a,** Microscope image of the few-layer $Bi_2Sr_2CaCu_2O_{8+x}$ (BSCCO) sample encapsulated with hexagonal boron nitride (hBN). **b,** AFM measurement of the sample thickness of d = 28 nm, which corresponds to approximately 8-9 unit cells. **c,** Schematic of the experimental layer stack. The NIR pump triggers the THz emission from the spintronic emitter through the sapphire substrate. The distributed Bragg reflector (DBR) reflects the remaining non-absorbed pump in the backwards direction and transmits the THz field. The sketch shows two configurations: one where the THz field is generated beneath the sample (top), and one where it is generated away from the sample for a reference measurement (bottom). **d,** Normalized electric fields measured by electro-optic sampling with the sample (top, $E_{sample}$) and without the sample for a reference (bottom, $E_{ref}$). Both measurements are normalized to the reference peak field. **e,** Amplitude of the fast Fourier transform of the electric fields shown in panel d. **f** and **g,** Phase and amplitude of the complex transmission coefficient, calculated by dividing the complex Fourier spectra of the sample and reference fields.

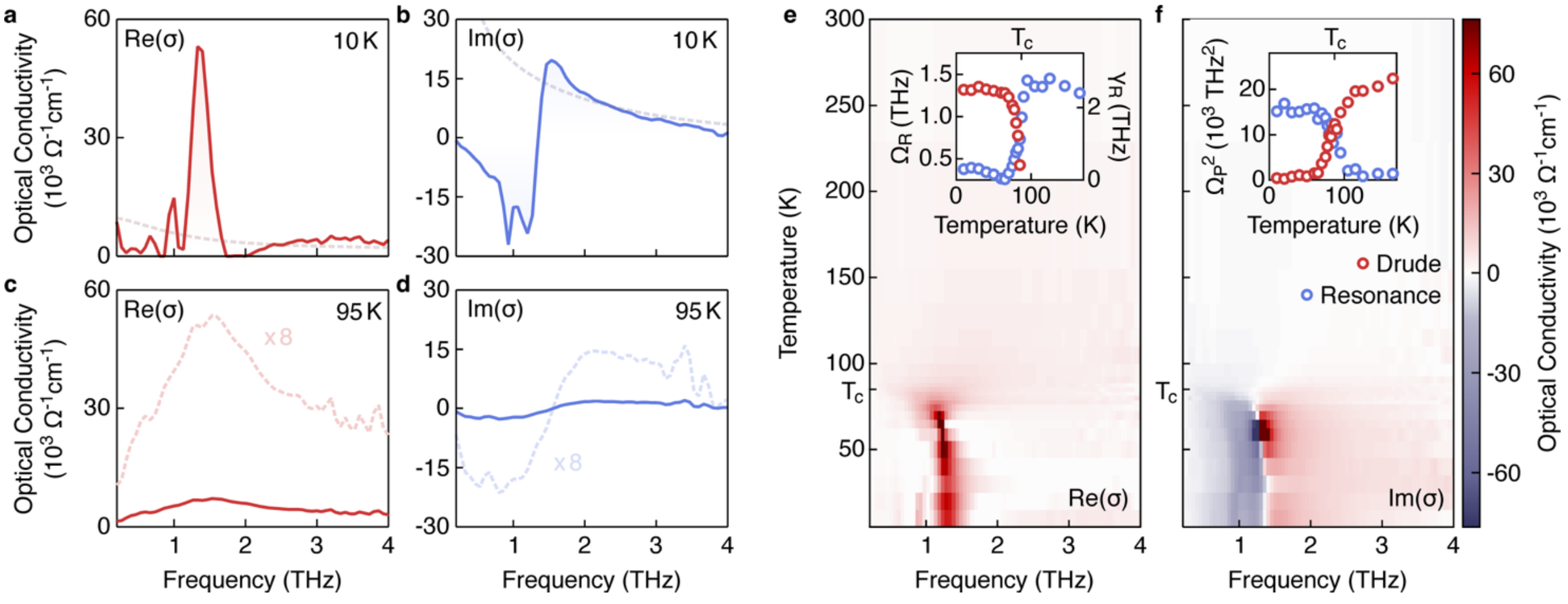

**Figure 2: Temperature dependence of the THz conductivity. a** and **b,** Real and imaginary parts of the THz conductivity at 10 K. The dashed lines are the THz conductivity of bulk $Bi_2Sr_2CaCu_2O_{8+x}$ adapted from Ref. 40. **c** and **d**, Real and imaginary parts of the THz conductivity above $T_c$ at 95 K. The dashed lines are the same data multiplied by eight. **e** and **f** show the temperature dependence of the real and imaginary parts of the THz conductivity, respectively. The inset in panel e shows the temperature dependence of the resonance frequency $\Omega_R$ (red circles) and its width $\gamma_R$ (blue circles) determined by tracking the maximum in the real part of the THz conductivity and its spectral width, respectively. The inset in panel f shows the temperature dependence of the amplitude $A \propto \Omega_P^2$ of the Drude contribution (red circles) and the resonance (blue circles) determined from the Drude-Smith-Lorentz analysis.

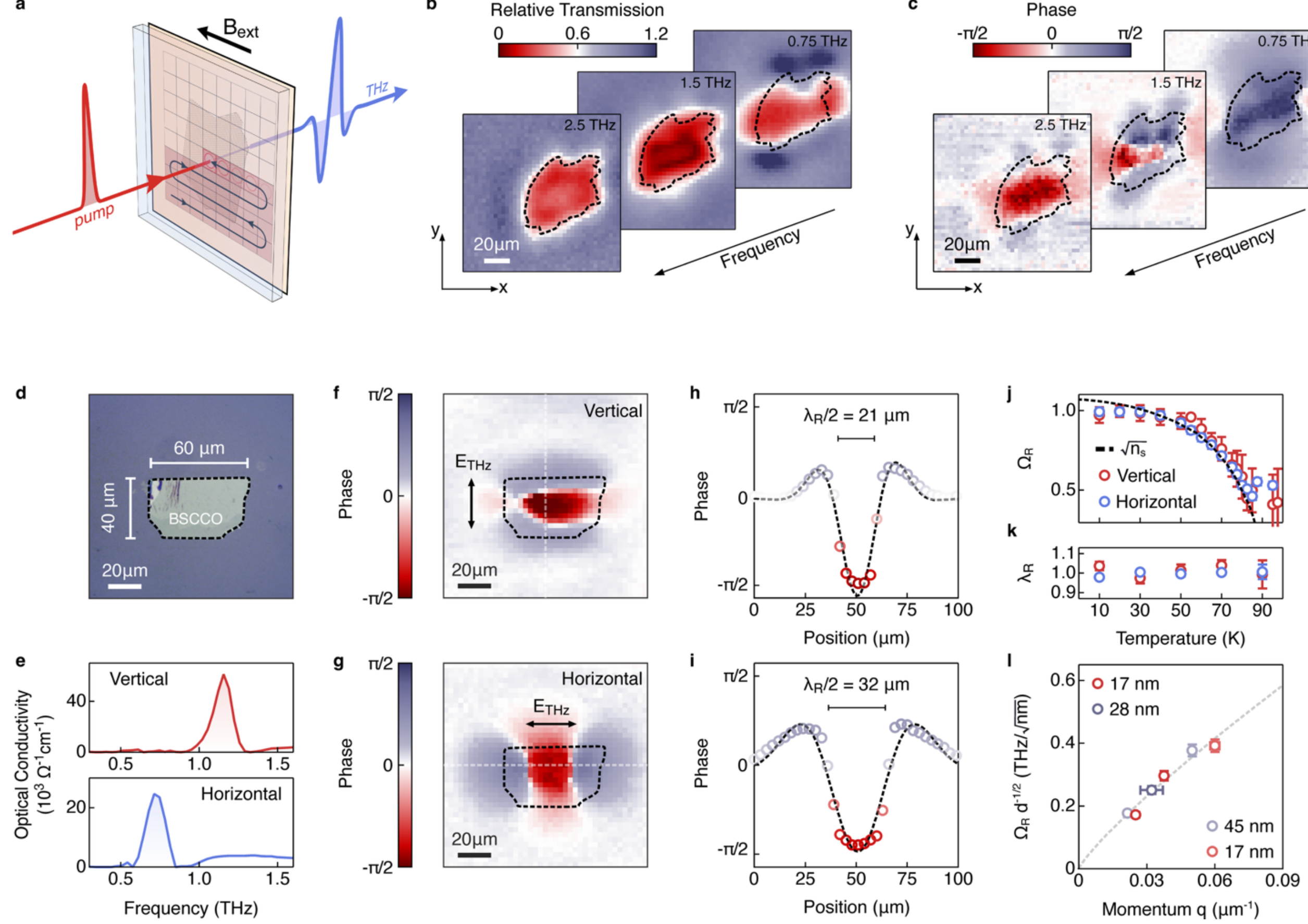

**Figure 3: THz hyperspectral imaging of $Bi_2Sr_2CaCu_2O_{8+x}$. a**, Schematic of the THz hyperspectral imaging setup. A pump pulse triggers local THz emission from the spintronic emitter, while the external magnetic field $B_{ext}$ controls the polarization direction of the emitted THz electric field. For imaging, the excitation beam is raster scanned across the emitter, and the spatially resolved THz field is detected via electro-optic sampling. **b** and **c**, Spatially resolved relative amplitude and phase of the THz field transmission coefficient with respect to the bare spintronic THz emitter at three frequencies. The dashed line indicates the outline of the BSCCO flake determined by optical microscopy (see Figure 1 a). **d**, Optical microscopy image of the second BSCCO flake. **e**, Real part of the THz conductivity at 10 K, measured at the center of the flake with vertical (top) and horizontal

(bottom) THz polarization. **f** and **g**, Spatially resolved phase of the THz transmission with vertical (panel f) and horizontal (panel g) THz polarization at the resonance frequency. The arrow indicates the THz polarization direction. **h** and **i**, Line cuts along the dashed lines in panel f and g, respectively. We determined the wavelength $\lambda_R$ of the spatial pattern through the full-width at half maximum of the negative peak. **j**, Normalized resonance frequency $\Omega_R/\Omega_R(T = 10\ K)$ as a function of temperature for vertical (red) and horizontal (blue) THz polarization, determined from the maximum in the real part of the THz conductivity and normalized to its value at 10 K (see Supplementary Data Figure S21). The dashed line is the normalized square root of the superfluid density adapted from Ref. 47. **k**, Wavelength $\lambda_R$ of the spatial pattern as a function of temperature for vertical (red) and horizontal (blue) THz polarization direction, extracted from Supplementary Data Figure S25. **l**, Normalized resonance frequency $\Omega_R/\sqrt{d}$ at 10 K measured on four different samples as a function of momentum q = 1/$\lambda_R$. The legend indicates the respective sample thickness, determined by atomic force microscopy (see Supplementary Data Figure S26). The dashed line is a weakly sub-linear fit to the data. Error bars represent the standard deviation σ propagated from the uncertainty level of a least squares fit.

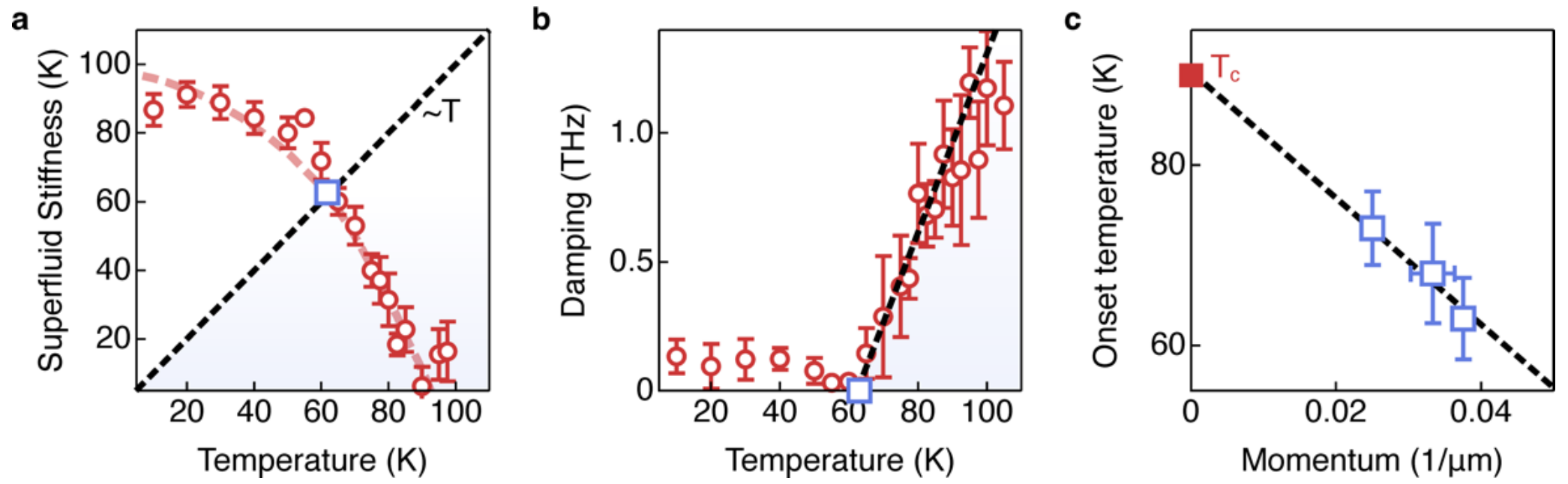

**Figure 4: Scale dependence of the superconducting transition. a**, Measured superfluid stiffness for vertical THz polarization (red circles) as a function of temperature. At low temperatures, the superfluid stiffness changes only slightly; once it becomes comparable to thermal fluctuations (dashed line, blue square), it drops abruptly and vanishes at $T_c$. The red dashed line is the normalized superfluid density adapted from Ref. 47 **b**, Damping rate of the superfluid plasmon as a function of temperature for vertical THz polarization. Above $T_{onset} \sim 62$ K, the damping rate sharply increases. The blue square indicates the temperature where the thermal fluctuations match the superfluid stiffness (panel **a**). **c**, Momentum dependence of the onset temperature extracted from the damping data in panel **b** (see also Supplementary Data Figure S28). The dashed line is a linear extrapolation from the data points (blue squares) to zero momentum, recovering the expected $T_c$ of our BSCCO sample. Error bars represent the standard deviation σ propagated from the uncertainty level of a least squares fit.

## Acknowledgement

We acknowledge the support from the US Department of Energy, Materials Science and Engineering Division, Office of Basic Energy Sciences (BES DMSE) (data taking, analysis and manuscript writing) and the EPiQS Initiative grant GBMF9459 (instrumentation) of the Gordon and Betty Moore Foundation. The theory work was supported in part by the European Research Council (ERC-2024-SyG-101167294; UnMySt), the Cluster of Excellence: Advanced Imaging of Matter (AIM), and Grupos Consolidados UPV/EHU (IT1249-19). We also acknowledge support from the Max Planck-New York City Center for Non-Equilibrium Quantum Phenomena. The Flatiron Institute is a division of the Simons Foundation. P.K. and X.C. acknowledge support from AFOSR (FA9550-25-1-0019)." A.v.H. acknowledges support from the Alexander-von-Humboldt Foundation. E.V.B. acknowledges funding from the European Union's Horizon Europe research and innovation program under the Maria Skłodowska-Curie grant agreement No 101106809. T.T. acknowledges support by the National Science Scholarship by the Agency for Science, Technology and Research (A*STAR), Singapore. M.Y. acknowledges support from the National Science Foundation MPS-Ascend Postdoctoral Research Fellowship under grant no. 2402151. K.T. acknowledges support from the NSF GRFP Fellowship Fabrication was carried out through the use of MIT.nano. Work at Brookhaven national laboratory is supported by US DOE (DE-SC0012704)

## Author contributions

A.v.H., C.A. and N.G. conceived the project. A.v.H. lead the experimental effort. A.v.H., T.T. and C.A. built the THz microscopy setup. T.T. developed the measurement software.

A.v.H. and T.T. performed the THz microscopy measurements with support from C.A.. A.v.H. and T.T. analyzed the experimental data with support from C.A.. B.L. and A.E.K. fabricated the spintronic emitter supervised by G.S.D.B.. M.Y. designed and deposited the distributed Bragg reflectors with support from A.v.H. M.Y. and J.C. fabricated the metallic maker structures used for the calibration measurements. X.C. and K.T. prepared the BSCCO samples supervised by P.K.. G.G. provided the crystals. J.P. carried out the AFM characterization. M.M. developed the theoretical description of the experiment with support from E.V.B supervised by A.R.. A.v.H. wrote the manuscript with support from T.T. and input from all authors. The project was supervised by N.G..

**Conflicts of interest**

The authors declare no competing interests.

**Data availability**

Datasets collected and/or analysed during the current study are available from the corresponding author upon request. Source data are provided with this paper.

**Code availability**

The codes used to generate the data for this study are available from the corresponding author upon request.

**Materials & Correspondence**

Correspondence to N. Gedik